\documentclass[twocolumn]{aastex631}
\shorttitle{Self-discharge by streaming cosmic rays}
\shortauthors{Yutaka Ohira}
\begin{document}

\title{Self-discharge by streaming cosmic rays}
\author[0000-0002-2387-0151]{Yutaka Ohira}
\affiliation{Department of Earth and Planetary Science, The University of Tokyo, \\
7-3-1 Hongo, Bunkyo-ku, Tokyo 113-0033, Japan}
\email{y.ohira@eps.s.u-tokyo.ac.jp}

\begin{abstract}
A new nonthermal phenomenon caused by streaming cosmic rays (CRs) in the universe is proposed. 
The streaming CRs drive the return current of thermal electrons to compensate for the CR current. 
Then, electric fields are induced by the resistivity of the return current. 
It is shown that the resistive electric fields can accelerate secondary electrons generated by the streaming CRs.
This is the self-discharge by streaming CRs. 
In this work, the self-discharge condition and the condition for runaway acceleration of  secondary electrons are presented. 
The self-discharge makes high-energy secondary electrons, resulting in enhancements of ionization and nonthermal emission including K$\alpha$ emission line of neutral irons. 
After the self-discharge, the return current of thermal electrons is replaced by the electric current of secondary electrons. 
Since some magnetic field generations and amplifications are driven by the return current of thermal electrons, the self-discharge can significantly influence them. 
\end{abstract}

\section{Introduction}
\label{sec:1}
Thanks to many studies about cosmic-ray (CR) interactions with neutral gases and plasmas, it is known that CRs play various roles in different environments. 
Since the energy of CRs is much higher than the energy scale of any astrophysical objects, CRs can ionize and heat gas in the Universe \citep{hayakawa61,fujita07,fujita11}.  
In addition, the CR interactions generate light nuclei (e.g. Li, Be, B)\citep{tatischeff18}, gamma rays \citep{hayakawa52}, and x rays \citep{tatischeff12}.  
Furthermore, the CR pressure is not negligible in some environments, so that CRs affect the dynamical evolution of astrophysical phenomena \citep{drury81,jubelgas08}. 
Since CRs consist mainly of protons with a positive charge, 
streaming CRs make an electric current, so that a return current of thermal electrons is driven to neutralize the CR current. 
The return current of thermal electrons in nonuniform systems is expected to generate seed magnetic fields in the early universe \citep{miniati11,ohira20,ohira21}. 
In the current universe, it is known that the electron return current drives the nonresonant (Bell) instability \citep{bell04}, amplifying magnetic field fluctuations, which is expected to be important for CR acceleration \citep{bell04} and propagation \citep{blasi15}.

Interactions of CRs with Earth's atmosphere provide high-energy particles in the inner Van Allen belt \citep{li17}. 
In addition, CRs generate high-energy secondary electrons in Earth's atmosphere. 
There are electric fields in thunderclouds of Earth's atmosphere, which is generated not by CRs, but by the charge separation due to friction \citep{takahashi78}. 
Low-energy electrons cannot be accelerated by the electric fields because the friction with the atmosphere is large. 
On the other hand, high-energy secondary electrons can be accelerated in thunderclouds and generate further secondary electrons, resulting in electron avalanche \citep{dwyer12}. 
This discharge induced by CRs in thunderclouds is expected to be the origin of some types of gamma-ray flares from thunderclouds \citep{dwyer12}.

Streaming CRs in the universe induce resistive electric fields by the Coulomb interaction of the return current of thermal electrons \citep{miniati11,silsbee20}. 
In this work, we first investigate a condition that secondary electrons generated by CRs are accelerated by the resistive electric field induced by streaming CRs themselves, that is, the self-discharge condition. 
This is a new discharge mechanism and a new acceleration mechanism for secondary electrons in the universe. 
Furthermore, we show that the return current of thermal electrons is replaced by the electric  current of high-energy secondary electrons after the self-discharge. 
Since the resistivity of high-energy secondary electrons is smaller than that of low-energy electrons, the resistive electric field becomes small after the self-discharge. 
Therefore, the self-discharge inhibits the magnetic field generation and amplification by the return current of thermal electrons. 
Moreover, the self-discharge enhances ionization and nonthermal emission by secondary electrons. 

\section{Self-discharge condition}
\label{sec:2}
We first consider thermal electrons, protons, hydrogen atoms, and streaming CRs to derive the discharge condition. 
In the proton rest frame, thermal electrons have a drift velocity, $V_{\rm e}=(n_{\rm CR}/n_{\rm e}) V_{\rm CR}$, to satisfy the current neutrality condition, where $n_{\rm CR}, n_{\rm e}$, and $V_{\rm CR}$ are the CR density, electron density, and the streaming velocity of CRs. 
The CR charge is assumed to be the positive elementary charge, $e$. 
The return current of thermal electrons feels the resistivity due to the Coulomb interaction or interaction with neutral hydrogens. 
For gases with the electron temperature of $T_{\rm e}=10~{\rm K}$, the Coulomb resistivity is dominant as long as the electron fraction is larger than $\sim 5\times 10^{-8}$. 
The time scale for the Coulomb scattering is 
\begin{equation}
t_{\rm C}=1.9\times10^{-1}~{\rm sec} \left(\frac{n_{\rm e}}{1~{\rm cm}^{-3}}\right)^{-1} \left(\frac{T_{\rm e}}{10~{\rm K}}\right)^{\frac{3}{2}}.
\label{eq:tc}
\end{equation}
Then, the resistive electric field before discharge is given by Ohm's law, 
\begin{eqnarray}
E_0 &=& -\frac{m_{\rm e}n_{\rm CR}V_{\rm CR}}{en_{\rm e}t_{\rm C}}   \label{eq:efield} \\
   &=& -9.5\times10^{-14}~{\rm V~cm^{-1}} \nonumber \\
   &&\times \left(\frac{V_{\rm CR}}{c}\right) \left(\frac{n_{\rm CR}}{10^{-9}~ {\rm cm}^{-3}}\right) \left(\frac{T_{\rm e}}{10~ {\rm K}}\right)^{-\frac{3}{2}} \nonumber ,
\end{eqnarray}
where $m_{\rm e}$ is the electron mass. 
The Coulomb resistive electric field induced by streaming CRs does not depend on the density of thermal particles but depends on the CR current and the electron temperature.

The typical energy of secondary electrons generated by the CR interaction with hydrogen atoms is $\varepsilon \sim10~{\rm eV}$ \citep{rudd92}. 
The secondary electrons lose their energy due to the excitation, ionization, and Coulomb  losses. 
If the energy gain from the electric field, $E$, is larger than the energy loss, the secondary electrons can be accelerated by the electric field, resulting in the discharge. 
Therefore, the discharge condition is given by
\begin{equation}
\frac{{\rm d}\varepsilon}{{\rm d}t} = -eEv+ \dot{\varepsilon}_{\rm loss} > 0,
\label{eq:dc1}
\end{equation}
where $v$ is the velocity of secondary electrons and $\dot{\varepsilon}_{\rm loss}$ is the energy loss rate. 
If the discharge condition is satisfied at $\varepsilon \sim10~{\rm eV}$, almost all secondary electrons are accelerated.

\begin{figure}
\epsscale{1.2}
\plotone{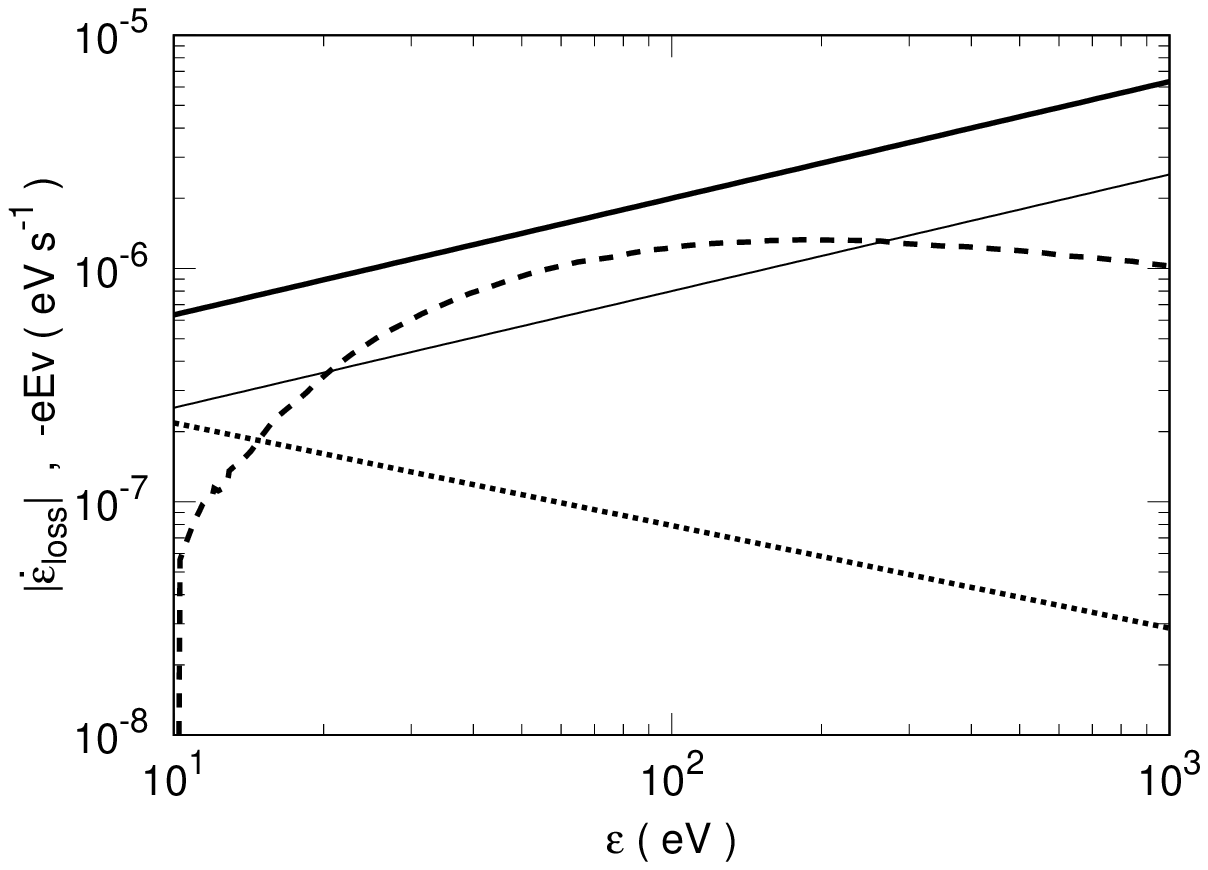}
\caption{Energy gain and loss rates for electrons. 
The dashed and dotted lines show the loss rates, $|\dot{\varepsilon}_{\rm loss}|$, for the excitation and ionization processes in $n_{\rm H}=1~{\rm cm}^{-3}$ \cite{dalgarno99}, and for the Coulomb interaction in $n_{\rm e}=3\times 10^{-3}~{\rm cm}^{-3}$ \cite{swartz71}, respectively. 
The thin and bold solid lines show the energy gain by the resistive electric field, $-eEv$, for cases below and above the runaway acceleration condition, Equation~(\ref{eq:rc}), respectively.}
\label{f1}
\end{figure}

Figure~\ref{f1} shows energy gain and loss rates as a function of the electron energy \citep{dalgarno99}. 
The dashed and dotted lines show loss rates for the excitation and ionization processes in $n_{\rm H}=1~{\rm cm}^{-3}$, and for the Coulomb interaction in $n_{\rm e}=3\times 10^{-3}~{\rm cm}^{-3}$. 
These energy loss rates linearly depend on hydrogen and electron densities.  
For an electron fraction of $n_{\rm e}/n_{\rm H} \le 8.3\times 10^{-4}$, excitation and ionization are the dominant loss process for electrons of $\varepsilon \ge 10~{\rm eV}$. 
Then, from Equation~(\ref{eq:efield}) and the energy loss rates at $\varepsilon=10~{\rm eV}$ of $\dot{\varepsilon}_{\rm loss}=-6\times 10^{-8} ~{\rm eV~s^{-1}}(n_{\rm H}/1~{\rm cm}^{-3})$, the self-discharge condition (\ref{eq:dc1}) is reduced to the following inequality for the CR flux,
\begin{equation}
n_{\rm CR}V_{\rm CR} > 3.2\times 10^{-1}~{\rm cm^{-2}~ s^{-1}} \left(\frac{T_{\rm e}}{10~{\rm K}}\right)^{\frac{3}{2}} \left(\frac{n_{\rm H}}{1~{\rm cm}^{-3}}\right).
\label{eq:dc2}
\end{equation}
For a lager electron fraction ($n_{\rm e}/n_{\rm H} \ge 8.3\times 10^{-4}$), the secondary electrons of $\varepsilon \sim 10~{\rm eV}$ lose their energy by the Coulomb loss. 
Then, the self-discharge condition is reduced to   
\begin{equation}
n_{\rm CR}V_{\rm CR} > 3.9\times 10^{2}~{\rm cm^{-2}~ s^{-1}} \left(\frac{T_{\rm e}}{10~{\rm K}}\right)^{\frac{3}{2}} \left(\frac{n_{\rm e}}{1~{\rm cm}^{-3}}\right),
\label{eq:dc3}
\end{equation}
where $\dot{\varepsilon}_{\rm loss}=-2\times 10^{-4}~{\rm eV~s^{-1}}(n_{\rm e}/1~{\rm cm}^{-3})(\varepsilon/1~{\rm eV})^{-0.44}$ is used for the Coulomb loss \citep{swartz71}. 
As one can see, the self-discharge condition is satisfied more easily in gases with lower temperatures and lower densities because the lower temperature makes the resistive electric field larger and the lower density makes the energy loss of the secondary electrons smaller.

\section{Acceleration of secondary electrons}
\label{sec3}
Once the self-discharge occurs, secondary electrons are accelerated by the electric field. 
If the electric field is not sufficiently large (see the thin solid line in Figure~\ref{f1}), the acceleration of secondary electrons stops at several $10~{\rm eV}$ because the energy loss due to the ionization and excitation becomes larger than the energy gain by the electric field. 
However, for larger electric fields, that is, larger CR fluxes (see the bold solid line in Figure~\ref{f1}), secondary electrons continue to be accelerated until the electric field disappears. 
The runaway acceleration condition is 
\begin{equation}
n_{\rm CR}V_{\rm CR} > 2.4~{\rm cm^{-2}~ s^{-1}} \left(\frac{T_{\rm e}}{10~{\rm K}}\right)^{\frac{3}{2}} \left(\frac{n_{\rm H}}{1~{\rm cm}^{-3}}\right).
\label{eq:rc}
\end{equation}

If both the discharge and runaway acceleration conditions are satisfied, the secondary electrons could be accelerated to very high energies. 
However, if the return current of the secondary electrons compensates for the CR current, the return current of thermal electrons and the resistive electric field become small. 
To estimate the maximum energy of secondary electrons, we have to understand the evolution of the electric field. 
In this work, we consider a simple one-zone system as the first step. 
The time evolution of the number density of secondary electrons is given by 
\begin{eqnarray}
n_{\rm 2nd} &=& n_{\rm CR} \left(\frac{t}{t_{\rm ion}}\right) + n_{\rm 2nd,0}~~,\label{eq:n2nd} \\
t_{\rm ion}&=&2\times10^8~{\rm sec} ~\left(\frac{n_{\rm H}}{1~{\rm cm}^{-3}}\right)^{-1},
\label{eq:tion}
\end{eqnarray}
where $n_{\rm 2nd,0}$ is the density of background secondary electrons that are generated by isotropic background Galactic CRs and $n_{\rm 2nd,0}\sim 10^{-10}-10^{-8}\ {\rm cm}^{-3}$ tipically \citep{ivlev21}, and $t_{\rm ion}$ is the ionization time scale of CRs with the energy of $1~{\rm GeV}$. The equation of motion for the secondary electron fluid is 
\begin{equation}
m_{\rm e}\frac{{\rm d}}{{\rm d}t}(n_{\rm 2nd}V_{\rm 2nd}) = -eEn_{\rm 2nd},
\label{eq:eom2nd}
\end{equation}
where $V_{\rm 2nd}$ is the mean velocity of secondary electrons and the resistive force is ignored here because the runaway acceleration condition is now satisfied. 
From the current neutrality condition, $n_{\rm CR}V_{\rm CR} = n_{\rm e}V_{\rm e} + n_{\rm 2nd}V_{\rm 2nd} $, the equation of motion for the thermal electron fluid is given by
\begin{equation}
m_{\rm e}\frac{{\rm d}}{{\rm d}t}(n_{\rm e}V_{\rm e}) = -m_{\rm e}\frac{{\rm d}}{{\rm d}t}(n_{\rm 2nd}V_{\rm 2nd}).
\label{eq:eome}
\end{equation}
From Equations (\ref{eq:n2nd}), (\ref{eq:eom2nd}), (\ref{eq:eome}), and Ohm's law, $E=-m_{\rm e}V_{\rm e}/et_{\rm C}$, we can obtain the following equation for the electric field, 
\begin{equation}
\frac{{\rm d}E}{{\rm d}t}= - \frac{n_{\rm CR}t+n_{\rm 2nd,0}t_{\rm ion}}{n_{\rm e}t_{\rm ion}t_{\rm C}} E.
\label{eq:dedt}
\end{equation}
The analytical solution is given by
\begin{equation}
E = E_0 \exp\left\{ - \left (\frac{t}{t_{\rm d,1}} \right) - \left( \frac{t}{t_{\rm d,2}} \right)^2   \right\} ,  
\label{eq:e}
\end{equation}
where $E_0$ is the initial electric field given by Equation~(\ref{eq:efield}), and the characteristic decay times, $t_{\rm d,1}$ and $t_{\rm d,2}$, are defined by 
\begin{eqnarray}
t_{\rm d,1} &=& \frac{n_{\rm e}}{n_{\rm 2nd,0}} t_{\rm C}  \label{eq:td0} \\
&=& 1.9 \times 10^{8}~{\rm sec} \left(\frac{n_{\rm 2nd,0}}{10^{-9}~{\rm cm}^{-3}} \right)^{-1} \left(\frac{T_{\rm e}}{10~{\rm K}} \right)^{\frac{3}{2}}, \nonumber \\
t_{\rm d,2}    &=& \sqrt{\frac{2n_{\rm e}t_{\rm ion} t_{\rm C} }{n_{\rm CR}}} \label{eq:td} \\
&=& 2.8\times 10^8~ {\rm sec} \left(\frac{n_{\rm CR}}{10^{-9}~{\rm cm}^{-3}} \right)^{-\frac{1}{2}}\left(\frac{n_{\rm H}}{1~{\rm cm}^{-3}} \right)^{-\frac{1}{2}} \left(\frac{T_{\rm e}}{10~{\rm K}} \right)^{\frac{3}{4}}. \nonumber 
\end{eqnarray}
After the decay time ($t>\min(t_{\rm d,1},t_{\rm d,2})$), the electric current of secondary electrons becomes sufficiently large to compensate for the CR current. 
Then, to satisfy the current neutrality condition, the return current of thermal electrons becomes small, so that the resistive electric field becomes small.

The maximum energy of accelerated secondary electrons can be estimated by solving the equation of motion for the first secondary electron, ${\rm d}p/{\rm d}t= - eE(t)$. 
From Equation (\ref{eq:e}), the maximum momentum is given by
\begin{equation}
p_{\rm max} = p_0 - \frac{\sqrt{\pi}}{2}eE_0 t_{\rm d,2} {\rm erfcx}(t_{\rm d,2}/2t_{\rm d,1}), \label{eq:pmax}
\end{equation}
where $p_0$ is the initial momentum of secondary electrons, and ${\rm erfcx}(x)=(2/\sqrt{\pi})e^{x^2}\int_x^{\infty}e^{-t^2}dt$ is the scaled complementary error function.

In dense molecular clouds, many secondary electrons are generated by streaming CRs before background secondary electrons are accelerated by the resistive electric field. 
Then, the effect of the background secondary electrons is negligible, resulting in $t_{\rm d,1} > t_{\rm d,2}$. In this case, the maximum momentum is given by
\begin{eqnarray}
p_{\rm max} &\approx& p_0 - \frac{\sqrt{\pi}}{2}eE_0 t_{\rm d,2} \nonumber \\
&\approx& m_{\rm e}V_{\rm CR} \sqrt{\frac{\pi n_{\rm CR}t_{\rm ion} }{2n_{\rm e} t_{\rm C}}}\nonumber \\
                    &\approx& 1.3~ m_{\rm e}c \left(\frac{V_{\rm CR}}{c}\right) \left(\frac{n_{\rm CR}/n_{\rm H}}{10^{-9}}\right)^{\frac{1}{2}}  \left(\frac{T_{\rm e}}{10~{\rm K}}\right)^{-\frac{3}{4}} .
\label{eq:pmax2}
\end{eqnarray}

If the background CR density is large, the number density of background secondary electrons becomes large. Moreover, the resistive electric field is stronger in lower temperature regions. 
Then, the electric current of background secondary electrons quickly cancels the CR current before new secondary electrons are generated by streaming CRs, resulting in $t_{\rm d,1} < t_{\rm d,2}$. 
In this case, the maximum momentum is given by
\begin{eqnarray}
p_{\rm max} &\approx& p_0 - eE_0 t_{\rm d,1} \nonumber \\
&\approx& m_{\rm e}c\left(\frac{V_{\rm CR}}{c}\right) \left(\frac{n_{\rm CR}}{n_{\rm 2nd,0}} \right).
\label{eq:pmax1}
\end{eqnarray}

In the following, we consider two specific environments in the universe as examples. 

\section{Application to molecular clouds in the vicinity of supernova remnants in the current universe}
\label{sec4}
In the current universe, supernova remnants (SNRs) are believed to inject about $U_{\rm CR}=10^{50}~{\rm erg}$ in CRs. 
Since the size of SNRs, $R_{\rm SNR}$, is typically from a few pc to 10 pc, the number density of GeV CRs is expected to be $n_{\rm CR}\sim 4.4\times 10^{-6}~{\rm cm}^{-3} (U_{\rm CR}/10^{50}~{\rm erg})(R_{\rm SNR}/5~{\rm pc})^{-3}$ around SNRs. 
Gamma-ray observations show that CRs interact with molecular clouds in the vicinity of SNRs \citep{ohira11,ackermann13}, supporting $U_{\rm CR}=10^{50}~{\rm erg}$. 
For a molecular cloud with $T_{\rm e} = 10~{\rm K}, n_{\rm H} = 10^3~{\rm cm}^{-3}, n_{\rm e}=1~{\rm cm}^{-3}$, the discharge (Equation~(\ref{eq:dc3})) and runaway acceleration (Equation~ (\ref{eq:rc})) conditions are satisfied when the drift velocity of CRs, $V_{\rm CR}$, is larger than $3\times 10^{-3}~ c$ and $1.8\times 10^{-2}~c$, respectively. 
The drift velocity of CRs is at least on the order of the shock velocity of SNRs, which is $10^{-3} ~c$ for middle-aged SNRs and $10^{-2}~c$ for young SNRs. 
Therefore, both conditions are expected to be satisfied marginally. 
Escaping CRs have the drift velocity faster than the shock velocity of SNRs, although the CR density becomes smaller than the above estimate. 
The dilution of CR density depends on the diffusion coefficient around SNRs and the shape of magnetic field lines, which are open problems and actively investigated. 
If any signatures of the self-discharge are observed, we can obtain information about the CR diffusion. 

For the above parameter set, $t_{\rm d,1}$ is larger than $t_{\rm d,2}$. 
From Equation~(\ref{eq:pmax2}), the maximum momentum of runaway secondary electrons becomes $p_{\rm max} = 2.7 ~m_{\rm e} c (V_{\rm CR}/c)$. 
Therefore, the accelerated secondary electrons due to the self-discharge could enhance MeV gamma-ray and 6.4 keV Fe line productions, and the ionization rate in molecular clouds.

\section{Application to the first CRs in the early universe}
\label{sec5}
According to our recent study, the first CRs are accelerated by SNRs of the first stars at the redshift of $z\approx 20$ \citep{ohira19}. 
The mean number density of the first CRs is estimated to be $n_{\rm CR}= 3\times 10^{-14}~{\rm cm}^{-3}$ at $z\approx 20$. 
The drift velocity is almost the speed of light because there are no or very weak magnetic fields in the early universe, so that the mean CR flux is $10^{-3}~{\rm cm^{-2}~s^{-1}}$. 
On the other hand, the hydrogen and electron number densities and temperature are $n_{\rm H}\sim 10^{-3}~{\rm cm}^{-3}, n_{\rm e}\sim 10^{-7}~{\rm cm}^{-3}$ and $T_{\rm e} \sim 1~{\rm K}$ at $z\approx 20$. 
Therefore, both discharge (Equation~(\ref{eq:dc2})) and runaway acceleration (Equation~(\ref{eq:rc})) conditions are satisfied. 
Since there are no background isotropic CRs in the early universe, from Equation~(\ref{eq:pmax2}) ($t_{\rm d,1} > t_{\rm d,2}$), the maximum momentum of accelerated secondary electrons becomes $p_{\rm max} = 1.7~m_{\rm e} c (V_{\rm CR}/c)$. 
The accelerated secondary electrons could heat and ionize gases. 
In addition, the return current of thermal electrons quickly decays in the time scale of $t_{\rm d,2}\sim 6.7~{\rm kyr}$. 
Therefore, some generation mechanisms of magnetic fields by the return current of thermal electrons \citep{miniati11,ohira20} can work not for the cosmological timescale, but only for $t_{\rm d}$. 
However, inhomogeneity of the return current of secondary electrons can generate magnetic fields \citep{ohira20}. 
Moreover, the Biermann battery driven by the streaming CRs \citep{ohira21} can work for the cosmological timescale by the gradient of the electron pressure, which is generated during the decay time, $t_{\rm d}$. 
Either way, the self-discharge significantly influences the magnetic field generations by streaming CRs.

\section{Discussion}
\label{sec6}
We have compared the energy loss and acceleration by the resistive electric field at the energy of $10~{\rm eV}$, which is the typical energy of secondary electrons, to understand whether the self-discharge occurs or not. 
However, secondary electrons with higher energies are sometimes generated by CRs, where the production rate is proportional to $\varepsilon^{-1}$ \citep{rudd92}. 
The energy loss rate becomes small for higher energy secondary electrons. 
Hence, the self-discharge condition can be satisfied even though the CR flux is below the critical values given by Equations~(\ref{eq:dc2}) and (\ref{eq:dc3}), but the event rate of the discharge becomes smaller. 
It should be noted that secondary electrons can be generated in fully ionized plasmas by the Coulomb scattering and nuclear interactions. 
In particular, the production rate of the secondary electrons by the Coulomb scattering is almost  the same as that of ionization \citep{abraham66}. 
Therefore, the self-discharge by streaming CRs can occur in various regions.

In Section~\ref{sec3}, we have investigated the evolution of the resistive electric field and the return current of thermal electrons when the runaway acceleration condition is satisfied. 
Even though it is not satisfied, as long as the discharge condition is satisfied, secondary electrons with energies of several 10 eV continue to be generated until the electric current of secondary electrons compensates for the CR current. 
In such a case, the number of the secondary electrons would be much larger than that of CRs, so that a large ionization rate would be expected. 
In the energy scale of several 10 eV, we need Monte Carlo simulations because the continuous slowing-down approximation is not applicable for electrons \citep{diniz18}. 
Furthermore, we have applied the one-zone system as the first step. 
In order to accurately evaluate the ionization rate and high energy nonthermal emissions including 6.4 keV Fe line from accelerated secondary electrons, we have to perform more rigorous calculations that consider the spectrum of secondary electrons, stochastic behaviors, and spatial structures, which will be addressed in future work.

Furthermore, in Section~\ref{sec3}, we implicitly assumed that the CR current suddenly increases, but in reality it takes a finite injection time, $t_{\rm inj}$. 
If the injection time is shorter than the decay time of the electric field, our implicit assumption is valid. 
However, if the injection time becomes longer than the decay time ($t_{\rm inj}>\min(t_{\rm d,1},t_{\rm d,2})$), before the CR flux reaches the critical value for the runaway acceleration, the resistive electric field is reduced by acceleration of secondary electrons.  
Then, the resistive electric field cannot reach the critical value for the runaway acceleration even though the CR flux reaches the critical flux given by Equation (\ref{eq:rc}). 
In this case, the runaway acceleration does not occur but many secondary electrons with energies of several 10 eV are generated. 
For molecular clouds in the vicinity of SNRs, the decay time of electric field is $t_{\rm d,2}\sim 10^5~{\rm sec}$ for the parameter set in Section~\ref{sec4}. 
The minimum timescale of the CR injection is the acceleration time of GeV CRs,  
which is $t_{\rm acc} \sim10^4 ~{\rm sec} (V_{\rm sh}/3000~{\rm km~s^{-1}})^{-2} (B/100~{\rm \mu G})^{-1}$ for acceleration at parallel shocks \citep{drury83}, and $t_{\rm acc} \sim10^4 ~{\rm sec} (V_{\rm sh}/3000~{\rm km~s^{-1}})^{-1} (B/1~{\rm \mu G})^{-1}$ for acceleration at perpendicular shocks \citep{jokipii87,kamijima20}. 
$V_{\rm sh}$ and $B$ are the shock velocity and magnetic field strength, respectively. 
On the other hand, the longest timescale of the CR injection is the diffusion time, 
$t_{\rm diff}=L^2/D= 10^{10}~{\rm sec}(L/5~{\rm pc})^2(D/10^{28}~ {\rm cm^2~s^{-1}})^{-1}$, where $L$ and $D$ are the distance from the CR source and the diffusion coefficient of CRs. 
If the CR injection time is comparable to the acceleration time, $t_{\rm acc}$, the runaway acceleration can occur in the molecular clouds, but cannot occur for $t_{\rm inj} \sim t_{\rm diff}$. 
In the early universe, the decay time of the resistive electric field is $t_{\rm d,2}\sim 10^{11}~{\rm sec}$  (see Section~\ref{sec5}), which is larger than the acceleration and propagation times of CRs. 
Therefore, the runaway acceleration can occur in the early universe. 
To understand more accurately whether the runway acceleration occurs or not, 
we need to study the self-discharge process more specifically by providing specific density structures. It should be noted that the above argument is based on the simple one-zone system that we considered in Section \ref{sec3}. 
In reality, the gas density is not uniform, especially in the edge of molecular clouds. 
We should investigate the self-discharge process in the nonuniform system in the future.

We have shown that the return current of thermal electrons is replaced by the current of secondary electrons once the self-discharge occurs. 
The return current of thermal electrons drives the nonresonant (Bell) instability in weak magnetic field \citep{bell04}, which amplifies magnetic field fluctuations. 
There are many studies about the Bell instability \citep{reville08,niemiec08,riquelme09,ohira09,marret15} because the Bell instability is expected to have an important role on CR acceleration and propagation. 
In principle, the current of secondary electrons can drive the nonresonant instability, but the growth rate and the nonlinear evolution would be modified. 
Furthermore, secondary electrons would be scattered by the magnetic field fluctuations amplified by the Bell instability, which would enhance the anomalous resistivity of the secondary electrons and the resistive electric field. 
Therefore, the self-discharge would also be influenced by the Bell instability.

\section{Summary}
\label{sec6}
We have shown that secondary electrons generated by streaming CRs can be accelerated by the resistive electric field that is induced by the streaming CRs themselves, that is, the self-discharge is driven by the streaming CRs. 
The self-discharge and runaway acceleration conditions have been derived. 
After the self-discharge, the return current of thermal electrons is replaced by the electric current of the secondary electrons. 
Then, the resistive electric field decays because the resistivity of the secondary electrons is much smaller than that of thermal electrons. 
Since some magnetic field generations and amplifications are driven by the return current of thermal electrons, the self-discharge significantly influences them. 
If the runaway acceleration condition is satisfied, the secondary electrons can be accelerated to mildly relativistic energies. 
As a result of the self-discharge, the ionization, nonthermal emissions, and 6.4 keV line of neutral Fe are enhanced by the accelerated secondary electrons.

\begin{acknowledgments}
We are grateful to the referee for valuable comments that improved the paper. 
We also thank Diniz Gabriel, Teruaki Enoto, Kazuhiro Nakazawa, and Yuuki Wada for informative  discussions on discharge in thunderclouds. This work is supported by JSPS KAKENHI Grant Number JP19H01893, and by Leading Initiative for Excellent Young Researchers, MEXT, Japan. 
\end{acknowledgments}

\end{document}